\documentclass[twocolumn,numbers,amsmath]{revtex4}
\usepackage{graphicx}
\usepackage{dcolumn}
\usepackage{bm}
\usepackage{epsfig}

\usepackage{wrapfig}

\begin{document}

\title{Electronic momentum redistribution along bond axes of Fe and Ni}    
\author{W. H. Appelt$^{a,b}$, D. Benea$^{c,b}$, L. Chioncel$^{a,b}$}
 \affiliation{$^a$ Augsburg Center for Innovative Technologies, University of Augsburg, 
D-86135 Augsburg, Germany}
\affiliation{$^{b}$ Theoretical Physics III, Center for Electronic
Correlations and Magnetism, Institute of Physics, University of
Augsburg, D-86135 Augsburg, Germany}
\affiliation{$^{c}$Faculty of Physics, Babes-Bolyai University,
Kogalniceanustr 1, Ro-400084 Cluj-Napoca, Romania} 

\begin{abstract}
We discuss the momentum redistribution along nearest and next nearest neighbour bond axes of Fe and Ni, using the Shannon entropy formula. 
We find that within the combined Density Functional and Dynamical Mean Field Theory weight redistribution takes place towards lower momenta as a function of the local Coulomb parameter $U$.
This effect is more pronounced for Fe than Ni.
\end{abstract}

\maketitle

The interest in momentum space density studies increased together with the 
development of experimental techniques such as Compton scattering and other 
electron momentum spectroscopies. The Compton profile is closely related to 
the single particle momentum density of an interacting electronic system  
\cite{C85}. The single particle momentum density can also be seen as the 
diagonal elements of the one-particle reduced density matrix in the momentum 
space representation $n({\bf p},{\bf p}^\prime)$. 
The Fourier transformation is used to connect the  
$n({\bf p},{\bf p}^\prime)$ to its counterpart in position space 
$\rho({\bf r},{\bf r}^\prime)$. Although the two one-particle density 
matrices are connected, there is no  direct connection between the diagonal 
elements of the one-particle density matrices: the real space density
$\rho({\bf r})$ and the corresponding momentum density $n({\bf p})$.
%
The ground state density in position space is the fundamental 
quantity upon 
which the Density Functional Theory(DFT) is constructed~\cite{HK64,KS65} and 
the formulation of charge density functionals opened the path towards 
computational materials science. The fact that DFT is constructed as a real 
space energy functional is based on the famous theorem of Hohenberg and 
Kohn~\cite{HK64} which states that the total energy of a non-degenerate 
ground state is a unique functional of the real space density. The 
universality of the exchange and correlation functional is a consequence of 
the universality of the kinetic and interaction term in the Hamiltonian.  
Although Hohenberg-Kohn type of theorems have been formulated in momentum 
space they have not gained so much interest since the functional in momentum 
space was proven to be not universal~\cite{HE81}.

Several studies showed that $\rho({\bf r})$ and $n({\bf p})$ contain different 
chemical aspects about the system~\cite{C85,SE92}. In addition to the different
chemical information encoded in $\rho({\bf r})$ and $n({\bf p})$, information 
theory attempts to measure the information content, directly.
For the charge density the corresponding Shannon entropy~\cite{SH48} 
$S_{\bf \rho} = - \int \rho({\bf r}) \ln \ \rho({\bf r}) d{\bf r}$ has been 
studied also as a measure for the accuracy of basis sets~\cite{GS85,HS94}, 
electron correlations~\cite{HS94b} or geometrical changes~\cite{HS95}.
Information theoretical concepts have been already used in momentum space. 
In analogy to the coordinate representation, the Shannon information entropy 
in momentum space 
$S_{\bf n} = - \int n({\bf p}) \ln \ n({\bf p}) d{\bf p}$
was defined using a formally equivalent equation and replacing $\rho({\bf r})$
with the probability density function in momentum space $n({\bf p})$
\cite{GS85,HS94}. 
A generalization of the Heisenberg uncertainty relation has been derived by 
Bia\l ynicki-Birula and Mycielski \cite{BM75} and was shown that the sum 
$S_{\bf \rho} + S_{\bf n}$ cannot be decreased beyond a certain limit 
$3(1+ \ln \pi)$ in three dimensions~\cite{BM75}. From a informational 
theoretical point of view this lower bound is just a manifestation of 
the maximum information 
density in phase space. This bound underlines the interdependence between the 
real and momentum space: the uncertainty in predicting the momentum of a 
particle is not independent of the uncertainty to predict the position of the 
particle, but bounded by the maximum information content in phase space.
It is worth to mention that here we are not talking about uncertainty in the 
usual sense as in Heisenberg uncertainty principle. In contrast to Heisenberg 
the term ``uncertainty'' should be understood as the lack of information in a 
literal manner~\cite{BM75}. In this formulation the Shannon entropy in momentum 
space has also been the subject of many investigations~\cite{GS85,HS94,HS94b}, 
and its maximum was connected to a localized distribution in position space.

Motivated by the capability to compute momentum space quantities in the 
presence of electronic correlations we analyze the influence of the local 
Coulomb interaction on the electronic momentum redistribution along the bond axis in Fe and 
Ni within the framework of a combined DFT and Dynamical Mean Field Theory 
(DMFT)~\cite{MV89,GK96,KV04}.
We have previously addressed different chemical aspects of bonding in Fe and Ni
using the computed total and magnetic Compton profiles~\cite{BMC+12,CBE+14}. 
The comparison with the experimental data lead us to conclude that theoretical 
Magnetic Compton Profile (MCP) spectra are improved as local correlations are 
taken into account. 
%

The aim of this paper is to discuss the effects of strong Coulomb interactions
upon the bonding in Fe and Ni. Contrary to the usual DFT approach, in which 
bonding is studied with the help of the charge density in real space, here
we perform an analysis using momentum space quantities. In section
Sec.~\ref{sec:2} we formulate the Shannon information entropy as the 
uncertainty to measure a certain momentum in Fe and Ni along different 
bond directions using the Compton profile that serves as the probability 
density. In order to understand the connection between the Compton
profile and the directional entropy in subsection \ref{ssec:2a}
we study a $q-$Gaussian model which allows us to analyze the behaviour of
entropy as a function of the Compton profile line shape.  
In the subsequent subsection we analyze the results from the realistic 
LSDA+DMFT calculations on the directional entropies (Sec. \ref{ssec:2b}).
We conclude the present paper in section Sec.~\ref{sec:conc}.

\label{sec:2}
Within DFT off-diagonal parts of the one particle density matrix as well as two 
particle information (electronic interactions) are only indirectly embedded in 
the one particle density density $\rho({\bf r})$. A complete description 
of properties of a system may be obtained by investigating the one particle 
density matrix $\rho({\bf r},{\bf r'})$. Technically such studies can be 
performed only on finite systems~\cite{GM05}. However, within the DFT 
framework a better description of electronic interactions leads to an 
improved description of the ground state of the many-body system. In 
the same time DFT is a very natural way to understand the chemical 
bonding, since bonding effects are significant for the 
charge density of valence electrons.

Since within our approach it is possible to gain insight into the momentum 
distribution in different lattice directions our aim is to discuss the covalent bonding 
using momentum space quantities. The momentum density $n(\bf p)$ is generally 
defined as the probability of finding an electron anywhere in position space 
with a given momentum ${\bf p}$. Mathematically, it is the spin traced 
diagonal of the one-particle density matrix in momentum-space representation 
$n({\bf p},{\bf p'})$. To access this quantity we performed electronic 
structure calculations using the spin-polarized relativistic 
Korringa-Kohn-Rostoker (SPR-KKR) method in the atomic sphere approximation 
(ASA)~\cite{EKM11}. This method was recently extended to compute magnetic 
Compton profiles (MCPs)~\cite{SGST84,BME06,DB04}. In the case of magnetic 
sample the spin resolved momentum densities are computed from the corresponding
LSDA(+DMFT) Green's functions in momentum space as:
\begin{equation}\label{e7}
n_{m_s}(\vec p)={-\frac{1}{\pi} \int_{-\infty}^{E_F}
\Im G_{m_s}^{LDA(+DMFT)}(\vec p,\vec p,E)dE}
\end{equation}
where $m_s=\uparrow(\downarrow)$. The many-body effects for d-orbitals 
are described by means of DMFT~\cite{MV89,GK96,KV04}. The relativistic 
version of the so-called Spin-Polarized T-Matrix Fluctuation Exchange 
approximation~\cite{KL02, PKL05} impurity solver was used (T=400K). In our 
calculations we used values for the Coulomb parameter in the range of 
U = 1.4 to 2.3 eV and the Hund exchange-interaction J = 0.9 eV. The 
electron momentum densities are usually calculated for the principal 
directions ${\bf K}=[001], [110], [111]$ using an rectangular grid of 
200 points in each direction. The maximum value of the momentum in each 
direction is 8 $(\text{atomic units})$.

The directional Compton profile $J({\bf p}_z)$ represents a probability 
density function, termed also as one-dimensional momentum distribution.
It is defined for a particular direction in the momentum space ${\bf p}_z$ 
and is obtained by integrating the momentum density $n_{m_{s}}({\bf p})$
(Eq.\ref{e7}) over planes perpendicular to this direction: 
$J({\bf p}_z)=\int \text{Tr}_{m_{s}} n_{m_{s}}({\bf p}) dp_xdp_y$. Using the results of 
the combined density functional and DMFT for the directional Compton profiles, 
we propose to use the Shannon information entropy formula with the directional 
Compton profiles as probability density  
\begin{equation} \label{SJ}
S_{\bf K} = - \int J({\bf p}_z)\  \ln \left(J({\bf p}_z)/m \right) d{\bf p}_z \qquad {\bf K} || {\bf p}_z,
\end{equation}
with $m$ the invariant measure~\cite{jaynes}.
We call this quantity {\it directional entropy}.
A similar formula has been used to obtain approximations to the atomic Compton profiles given only the first few 
moments of the Compton profile~\cite{SG81,GS85}. 
We compare the values of the directional entropies Eq. (\ref{SJ}) computed 
along the [001], [110] and [111] direction of the fcc and bcc - structures 
of Ni and Fe, respectively. The directional entropy provides the uncertainty 
in predicting the momentum in a certain lattice direction and 
therefore may provide information about chemical bonding.

\subsection{Entropy formula for a $q-$Gaussian model of the Compton profile } 
\label{ssec:2a}
\begin{figure}[htp]
\vspace{0.5cm}
\begin{center}
\hspace*{0.2cm}
   \includegraphics[width=0.95\linewidth, clip=true]{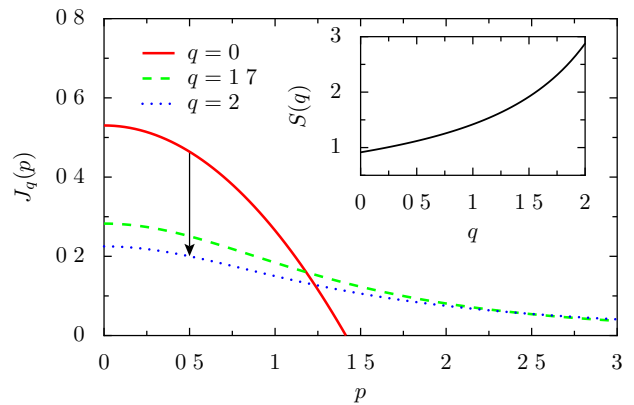}
\end{center}
\caption {\label{Fig:figure1} $q-$Gaussian probability distribution
taken as a simplified model Compton profile. Red solid line represents
the Compton profiles for the non-interacting homogeneous electron gas
(inverted parabola). Two other examples for $q=1.7$ (Dashed green) and 
$q=2.0$ (blue dotted) are plotted as function of $p$. The arrow indicates the direction of increasing $q$. 
Inset: Entropy computed using the Compton profiles as a function of $q$.}
\end{figure}

In order to clarify the connection between the directional Compton profile
and the corresponding entropy Eq. (\ref{SJ}), we discuss a simplified model
for the line shape of the Compton profile. The central question is how  
the shape of the profile is changing in the presence of strong electronic 
interactions and finite temperatures. In the most general case the line
shape is subject to a combined Lorentzian (excitations related) and Gaussian 
(Doppler related) broadening, also known as Voigt line-shape, which is just a convolution of the Gaussian and Lorentzian profile. In our simplified model we consider for the Compton profile a much simpler parametrization.
We use a generalization of the usual Gaussian distribution, called $q$-Gaussian:
\begin{equation}\label{qG}
J_q(p)=\frac{1}{C_q \sqrt{2} \sigma}\exp_q(-p^2/2\sigma^2),
\end{equation}
where the exponential function is replaced by its $q-$analog
\begin{equation}
        \exp_q(p)=(1+(1-q)p)^{1/(1-q)}
\end{equation}
and $C_q$ is the normalization factor. 
This has the advantage of describing also the Compton profile of the 
non-interacting electron gas ($q=0$), which is just an inverted parabola 
for momenta $p<p_F$, where $p_F$ is the Fermi momentum (zero otherwise):
$ J_{q=0}(p) \propto (p_F^2 -p^2)$.
In the limit $q\to1$ the usual Gaussian distribution with variance $\sigma^2$ 
is recovered, which has infinite support contrarily to the $q<1$ case. 
If we further increase $q$ beyond 1 the exponential tails of the Gaussian distribution turn into power law tails.
The case $q=2$ represents a model distribution of Compton profile that captures the limit of the one-bound state 
scatterer~\cite{KK77}.
For $0<q<5/3$ the variance of the $q$-Gaussian is given by $2\sigma^2/(5-3q)$. 
For $q-$values larger than 5/3 the variance diverges, and the 
uncertainty for this kind of probability distributions cannot be defined based on the moments of the distributions,
motivating the need for a different definition of uncertainty~\cite{BM75}.
Since entropy is a measure of the total amount of information in a distribution and since uncertainty is just the lack of information it is very natural to define uncertainty with the use of entropy.

As one can see in Fig. \ref{Fig:figure1} the increase in $q$ leads to heavier 
tails which are connected with a shift of weight from the region of higher 
probability density to region of lower probability density. 
The uncertainty in the prediction of the momentum is therefore increased as 
a function of $q$, which can be seen as an increase in entropy $S$ (see inset Fig. \ref{Fig:figure1}).
The magnitude of entropy has no meaning since we are considering probability 
densities instead of a discrete probability distributions, so the 
information content is only defined up to an irrelevant offset due 
to the choice of probability measure. In the formula for entropy:
\begin{equation}\label{sq}
S_q = - \int J_q({p})\  \ln\left(J_q({p})/m\right)d{p}
\end{equation}
We chose the invariant measure $m=const$ for brevity. 
The lack of a general (comprehensive) invariant measures makes difficulties in 
quantitative statements about uncertainty,
however it is still possible to make a relative comparison between 
two probability distributions providing the same invariant measure $m$. 
The simplest choice is the homogeneous measure $m$ which can be interpreted 
as a uniform discretization mesh of the probability density. 
With this choice the entropy still depends on the mesh size $\delta p$, 
which can arise from any finite resolution (the experimental setup) or 
the momentum step size in numerical calculation.
In experiment or in numerics the resolution can in principle be non-homogeneous for several reasons, 
which implies that any entropy construction is always subjective to the amount of 
information we have about the system.
Our choice of the invariant measure provides us with a definition of 
directional entropy so that the integrand in Eq. \ref{sq} is always positive. 

The analysis of the above model shows that increasing the $q$ parameter 
tails spread out towards higher moments and the entropy is increasing.
Conversely increasing entropy can be understood as weight redistributions that overall flattens the probability density. 

\label{ssec:2b}
In order to discuss chemical effects along the bonds in Fe and Ni we use
the directional entropy formula Eq. (\ref{SJ}) in which directionality 
enters through the Compton profile taken along the principal directions.
The later is computed from the momentum space total (spin-traced)
one-particle density matrix using LSDA and LSDA+DMFT Green's function
Eq. (\ref{e7}). 

\begin{figure}[htp]
\vspace{0.5cm}
\begin{center}
\hspace*{0.2cm}
   \includegraphics[width=0.95\linewidth, clip=true]{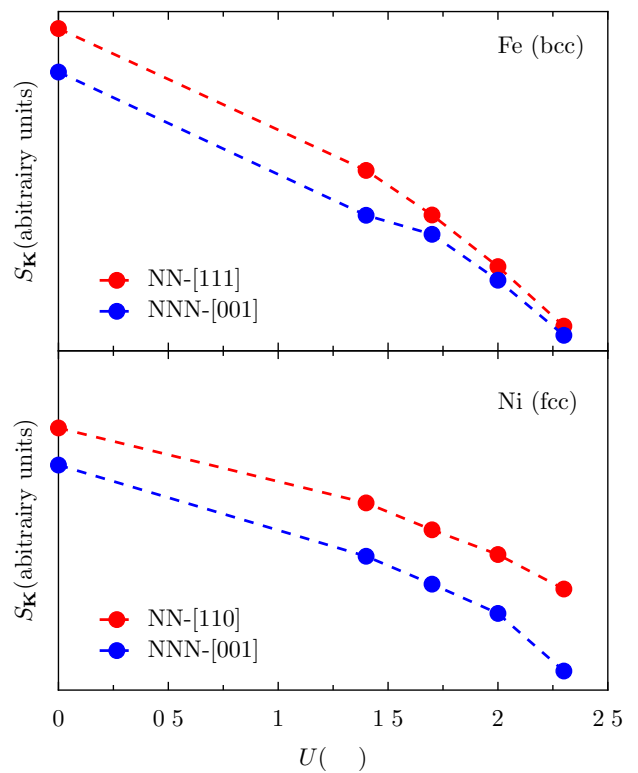}
\end{center}
\caption {\label{Fig:figure2} Directional Entropy for Fe/Ni 
(upper/lower pannel)  along the nearest neighbour (NN-) and 
next nearest neighbour (NNN - ) directions as a function
of U, fixed J=0.9 and T=400K.}
\end{figure}
Fig. \ref{Fig:figure2} shows the directional information entropy of Fe and 
Ni along the nearest neighbour and next nearest neighbours for different  
values of U. The LSDA values represent the results for U=0 (absence of 
local Coulomb repulsion). Including local but dynamic electronic 
correlations captured by DMFT, we see that the values of the directional
Shannon entropy decrease along all directions. A similar color coding was 
used for the nearest (NN - red) and next nearest (NNN - black) neighbours.
Given the geometry of the lattice NN and NNN bonding is realized along
different directions as seen in the legend of Fig.\ref{Fig:figure2}.
One can see that shorter bond lengths have larger entropies, and the U
dependence show a larger slope for Fe in comparison to Ni. The analysis of the entropy data suggests that
for increasing $U$ it is less likely to find electrons with nonzero 
momentum-component in a specific bond direction. 
Our findings agree with the calculation of the second moment $\langle p^2 \rangle$ of the Compton-profile~\cite{CBE+14}. 
We have interpreted the decreasing in kinetic 
energy as a function of $U$ as a shift of the weight of the momentum 
distribution towards zero momentum.
Therefore the Coulomb repulsion leads to a decrease in the uncertainty of the electron momentum, which can be 
understood also as the slowing down of the electrons.


\label{sec:conc}
In a simple valence electrons counting picture for bcc-Fe 8 bonds share 7 d-electrons, while fcc-Ni 12 bonds share 9 electrons. 
Therefore, Fe bonds are said to be more local then Ni bonds.
Electrons in open d-shell-systems are believed to interact strongly.
Strong interactions are modelled by a local Coulomb interaction parameter $U$, acting on the d-orbitals manifold. 
Model and realistic electronic structure calculations showed that for systems with narrow bands the effect of $U$ is to localize the valence electrons around the atoms, such that metallic conduction is no longer possible, so the system experience a localization of electrons through correlation effects~\cite{GK96,KV04}.
In our calculations for Fe and Ni we take correlation effects into account 
by means of LDA+DMFT and study momentum space quantities. Both Fe and Ni have
larger valence bandwidth than the realistic parameter for the 
average Coulomb interaction, therefore the lower- and upper-Hubbard bands are 
not present~\cite{KP10,AB12,LP11}. Although no strong localization is expected,
the question still remains to what extend the Fe/Ni electrons per bond localizes because of $U$ and how they compare.

In this paper we analyzed electronic properties from the one particle 
density matrix in momentum space within the information theoretical 
framework. In such a framework one defines a measure of information content 
or uncertainty. The most commonly used measure is the Shannon entropy, for 
which we proposed a formula that includes the directional Compton profiles. 
The directional entropy is a functional of the distribution of the momentum 
component in a certain direction ${\bf K}$. 
The Compton profile can be computed including electronic correlation within 
DMFT, therefore  we are able to consider electronic interactions consistently 
beyond the mean-field approximation and study their effect upon the chemical 
bonds in Fe and Ni. 
Our main result is that the probability of finding electrons with high momenta along bond axes is decreased in favour of low momenta as a function of $U$.

A possible consequence of the redistribution are briefly discussed below:
Fe and Ni have a metallic bonding with covalent d-d contribution. 
The covalent chemical bond is usually interpreted as electronic charge 
accumulation between nuclear centers. 
It is a dominant electrostatic approach and within DFT this effect is encoded into the diagonal of the real space density matrix $\rho({\bf r})$. 
Dynamical effects are usually neglected withing plain DFT.
Our numerical results suggest that the inclusion of local correlations within DMFT affects momentum distribution and therefore the covalent bonding.

Financial support of the
Deutsche Forschungsgemeinschaft through FOR 1346, the DAAD and the 
CNCS - UEFISCDI (project number PN-II-ID-PCE-2012-4-0470) is gratefully 
acknowledged. We would like to thank J. Min\`ar and Hubert Ebert for 
a fruitful collaboration.
    

\bibliography{paper}

\end{document}